\documentclass[%
 reprint,
 amsmath,amssymb,
 aps,
]{revtex4-2}

\usepackage{graphicx}
\usepackage{dcolumn}
\usepackage{bm}
\usepackage{xcolor}
\usepackage{hyperref}
\usepackage{braket}

\begin{document}

\preprint{APS/123-QED}

\title{Complete three-dimensional vector polarimetry with a Rydberg atom rf electrometer}

\author{Peter K. Elgee}
\email{peter.k.elgee.ctr@army.mil}
\author{Kevin C. Cox}
\author{Joshua C. Hill}
\author{Paul D. Kunz}
\author{David H. Meyer}
\affiliation{%
 DEVCOM Army Research Laboratory, 2800 Powder Mill Rd, Adelphi MD 20783, USA
}%
\date{\today}

\begin{abstract}
Radio frequency (rf) receivers using Rydberg atoms offer appealing features over classical sensors, such as their size, frequency tuning range, and lack of field absorption.
In this work, we extend the application space by demonstrating a Rydberg atom rf polarimeter.
Using rf heterodyne with three independent and orthogonal local oscillators, we are able to extract the polarization ellipse of the field in three dimensions, in addition to the total amplitude of the field.
We use the relative phases and amplitudes of the three generated heterodyne signals to measure the field amplitudes and phases along three cardinal axes giving the full three-dimensional polarization.
We demonstrate this polarization measurement for incoming fields at different angles around the sensor.
Lastly, we investigate the reception of data symbols encoded in the horizontal and vertical signal field polarizations and the phase between them.
Our measurements yield an amplitude noise of 57~$\mu$V/m/$\sqrt{\text{Hz}}$ for horizontal polarization, 66~$\mu$V/m/$\sqrt{\text{Hz}}$ for vertical polarization, and a standard deviation of 0.094 rad in the phase between the field components.

\end{abstract}

\maketitle

\section{\label{sec:Introduction} Introduction}
Many electric field sensors, both quantum and classical, provide the amplitude and phase of a field along some axis.
However, full characterization of the field at a point in space requires knowledge of the amplitude and phase in all three dimensions.
We demonstrate that this type of three-dimensional characterization is possible with Rydberg atom-based radio frequency (rf) receivers.
These receivers offer the promise of unique capabilities over classical antennas, notably their large tunability \cite{Downes20, Jau20, Meyer21}, accurate non-absorptive sensing \cite{Anderson21,Holloway14}, and small size \cite{Fan14}.
Apart from receivers, Rydberg atoms offer a clean platform for quantum electrodynamics \cite{ Sayrin11, Nogrette14, Ripka18, Kumar23, Evered23, Borowka24}, providing an opportunity to one day bridge classical rf reception and quantum information science.
With these advantages in mind, efforts have been made to extend the capabilities of these receivers, such as by measuring multiple tones simultaneously \cite{Meyer23}, tuning Rydberg levels to further expand the resonant sensing regime \cite{Berweger23}, or remotely interrogating a distant sensor \cite{Otto23}.
We show the capability of these Rydberg atom rf receivers to fully characterize the polarization of an incoming field, by measuring the polarization ellipse in three dimensions.
This is accomplished by leveraging multiple simultaneous rf heterodyne measurements \cite{Simons19, Jing20}, with three local oscillators (LOs) at orthogonal polarizations.
This expands the sensing capabilities to new areas, such as ellipsometry for material characterization, reception of polarization encoded information, and $k$-vector measurements for elliptical fields.

Full characterization of the electric field is an important and ongoing area of research and polarization measurements with Rydberg atom receivers have been demonstrated before \cite{Sedlacek13, Jiao17,Wang23, Berweger24}.
Initial experiments relied on asymmetric coupling strengths for the different field polarizations relative to the optical polarization axis \cite{Sedlacek13, Jiao17}.
Full characterization of these sensors relies on matching spectroscopy to first-principles models and the ability to change the optical polarization axis.
More recent work has measured the polarization projection relative to a single LO \cite{Wang23}.
This method only provides information about the polarization projection along one axis, is difficult to separate from overall amplitude changes, and does not provide uncoupled ellipticity information.
In contrast, our technique provides full three-dimensional polarization information, including ellipticity, and overall field amplitude from a single measurement.
In addition, previous $k$-vector measurements with Rydberg atoms have relied on the spatial extent of the interaction region being an appreciable portion of the carrier wavelength \cite{Robinson21}, while ellipticity measurements with our system could provide $k$-vector information with no such restriction.

In the following section we describe the operating principle of the system, followed by the experimental setup required for full reconstruction of the signal polarization.
We investigate angle and ellipticity measurements with the signal at a single $k$-vector, before extending the polarization space by altering the incoming $k$-vector.
Finally we demonstrate the ability for this system to receive symbols encoded in polarization and field amplitude.

\section{Methods}

\subsection{Operating principle}

Our technique operates in a thermal gas of rubidium and relies on rf heterodyne measurements.
In these heterodyne measurements a strong LO near resonance with a transition between Rydberg states provides improved sensitivity and acts as a phase reference for an incoming signal field\cite{Simons19, Jing20}.
The combined LO and signal fields cause an Autler-Townes splitting of the Rydberg states that oscillates at the difference frequency between the two fields.
This beat is read out through electromagnetically induced transparency (EIT) spectroscopy \cite{Mohapatra08}.
Our sensor uses three LOs with orthogonal linear polarizations (LO$_X$, LO$_Y$, LO$_Z$) as seen in Fig. \ref{fig:experimental_diagram}, and each is detuned from the signal frequency ($\omega_0$) by a different amount ($\delta_X$, $\delta_Y$, $\delta_Z$).
Thus the projection of the signal field along each polarization axis (Sig$_x$, Sig$_y$, Sig$_z$) can be extracted in Fourier space through its respective beat ($b_x$, $b_y$, $b_z$).
Throughout this work the above field quantities (LO$_i$, Sig$_i$, $b_i$) are taken to be complex, representing the amplitude and phase of each field in its rotating frame.
For instance, the electric field corresponding to LO$_X$ would be LO$_Xe^{i(\omega_0 + \delta_X)t} = |\text{LO}_X|e^{i\phi_X}e^{i(\omega_0 + \delta_X)t}$.

The three heterodyne measurements described above give a three-dimensional measurement of the polarization vector while simultaneously measuring the total signal field amplitude.
In addition, since we can simultaneously measure the signal beat with all LOs, we are able to extract the relative phase between all three orthogonal projections of the signal and thus fully characterize the polarization ellipse of the signal field.
This full polarization measurement indirectly provides information about the $k$-vector, as for any elliptical signal field the incoming $k$-vector is determined up to a sign, and for purely linear polarization the $k$-vector can be restricted to a plane.

Measurement of the field polarization with multiple simultaneous heterodyne measurements fundamentally depends on independent and symmetrically-sensitive detection from each polarization axis.
To quantify the independence of our orthogonal polarization measurements we can look at the dependence of the Autler-Townes splitting on rf fields polarized along $\hat{x}$, $\hat{y}$, and $\hat{z}$.
We can write the light-atom Hamiltonian restricted to the rf transition as
\begin{equation}\label{eq:hamiltonian}
\begin{split}
    H = &\ d_{-1}\left(\frac{1}{\sqrt{2}}E_x -\frac{i}{2}E_y + \frac{i}{2}E_z\right)\\
    &+ d_0\left(\frac{1}{\sqrt{2}}E_y+\frac{1}{\sqrt{2}}E_z\right)\\
    &+ d_{+1}\left(-\frac{1}{\sqrt{2}}E_x - \frac{i}{2}E_y + \frac{i}{2}E_z\right)
\end{split}
\end{equation}
where the dipole operator $d$ is represented in a spherical basis defined by the optical polarization axis, and the rf fields in the experimental basis of our LOs as shown in Fig. \ref{fig:experimental_diagram}.
In the simplest case of a signal field on-resonance with a $J = 0 \rightarrow 1$ transition at $\omega_0$, the atomic response is symmetric for all polarizations with dipole matrix element: $d \equiv \braket{1,-1|d_{-1}|0,0} = \braket{1,0|d_0|0,0} = \braket{1,1|d_1|0,0}$. 
Diagonalizing this Hamiltonian for the rf fields interacting with the Rydberg levels yields two sets of eigenvalues split by the Autler-Townes splitting.
In this case, the resulting splitting is the quadrature sum of the effective Rabi frequencies $\Omega_i = d |E_i|/\hbar$ along each axis: 
\begin{equation}\label{eq:at_splitting}
AT = \sqrt{\Omega_x^2 + \Omega_y^2 + \Omega_z^2},    
\end{equation}
Focusing on readout of the $\hat{x}$ projection we can write $E_x$ as a combination of LO$_x$ and Sig$_x$ as $E_x = \text{LO}_Xe^{i(\omega_0 + \delta_X)t} + \text{Sig}_xe^{i\omega_0t}$.
For small detunings, and signal amplitudes we can model the combined field as on resonance with an amplitude that oscillates at the corresponding beat frequency as \begin{equation}
    |E_x| \approx |\text{LO}_X|\left(1 + \left|\frac{\text{Sig}_x}{\text{LO}_X}\right|\cos(\delta_Xt + \phi_X - \phi_x)\right).
\end{equation}
Assuming the signal is small relative to the LOs ($\epsilon_i = |\text{Sig}_i/\text{LO}_i| \ll 1$) we can Taylor expand Eq. \ref{eq:at_splitting} around all three $\epsilon_i$ simultaneously.
Only terms with some component that oscillates at a frequency $\delta_x$ are relevant to our readout of the $\hat{x}$ projection of the field.
Thus, the terms to keep are those of the form $\epsilon_x^{2l + 1}\epsilon_y^{2m}\epsilon_z^{2n}$ for integers $l,n,m$, as odd orders of $\epsilon_x$ will have a term oscillating at $\delta_x$ and even orders of $\epsilon_y$ and $\epsilon_z$ will have DC terms.
This holds, provided the detunings are chosen to avoid overlap with harmonics and difference frequencies. 
The first such term with $(l,m,n) = (0,0,0)$ goes like
\begin{equation}
    \frac{|\text{LO}_X\text{Sig}_x|\cos(\delta_it + \phi_X - \phi_x)}{\sqrt{|\text{LO}_X|^2 + |\text{LO}_Y|^2 + |\text{LO}_Z|^2}}
\end{equation}
This term provides linear readout of Sig$_x$.
It is proportional to the relative amplitude of LO$_X$, and the phase is just the phase difference between the Sig$_x$ and LO$_X$ as expected.
Higher order terms interfere with simple independent readout of Sig$_x$, as they will depend on Sig$_y$ and Sig$_z$, but these terms are at least two orders of $\epsilon$ higher than the linear term.
We neglect these higher order terms in this work, but one could account for them through modeling.

The same principle applies when using higher $J$ levels, though the analysis becomes more complicated as the multiple $m_j$ sublevels have asymmetric couplings to different polarizations.
Fortunately, these coupling strengths are simply and fundamentally defined by the Clebsch-Gordan coefficients.
In contrast to Ref. \cite{Sedlacek13} where these asymmetries were required to read out the microwave polarization, in this work, a system that is symmetrically-sensitive to all polarizations is advantageous.
We will not fully quantify our polarization sensitivity symmetry here, but will indicate a few considerations that improve the symmetry in our system.
Firstly, we ensure that each sublevel of the initial Rydberg state has coupling with every polarization.
This is only true of a $J \rightarrow J+1$ transition, otherwise multiple sublevels of the initial Rydberg state, most clearly the stretched states, are not fully coupled \cite{Sedlacek13, You24}.
Even with coupling to all sublevels the strength of these couplings will depend on the Clebsch-Gordan coefficients.
Our linear optical polarizations help us with this effect in two ways.
First, we probe the $m_j$ and $-m_j$ sublevels equally in any basis, which provides symmetry between right and left-circularly polarized light along any axis.
Second, as the ground state has $J = 1/2$ we primarily optically probe the Rydberg sublevels with low $|m_J|$ quantized along the optical polarization \cite{Meyer23}, for these states the Clebsch-Gordan coefficients are more symmetric, as seen in Fig. \ref{fig:experimental_diagram} (b).
More detailed analysis of the variation in atomic response to polarization can be found in Refs. \cite{Sedlacek13, You24, Cloutman24}.
Given these asymmetries are based on the fundamental atomic physics of the system, there is an opportunity for precise and accurate modeling to correct for them.

\subsection{\label{sec:ExperimentalSetup} Experimental setup}

\begin{figure}[b]
\includegraphics[width = 3.375 in]{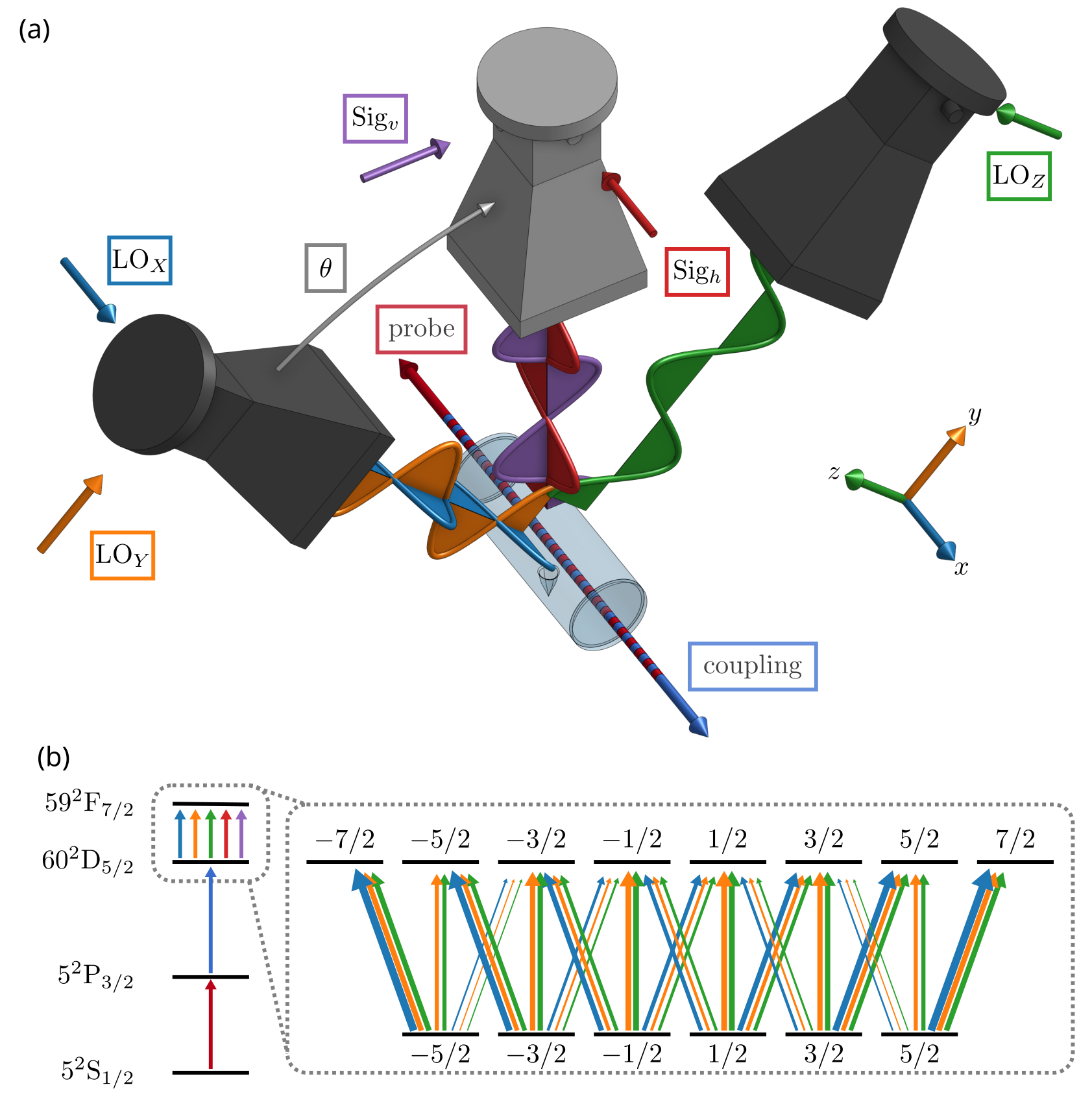}
\caption{\label{fig:experimental_diagram}
(a) A diagram of the experimental apparatus.
The vertically polarized probe and coupling lasers counter-propagate through the vapor cell.
Above the cell, two dual-polarization horns (dark grey) provide the three orthogonality polarized LOs positioned with their $k$-vectors orthogonal to the optical beams, and at 45$^\circ$ to the optical polarization.
Another horn (light grey) provides the signal field, and the polar angle $\theta$ of its $k$-vector relative to $\hat{z}$ can be manually varied.
(b) The level diagram for the experiment and an enlarged diagram of the $m_J$ sublevels of the Rydberg transition.
This is drawn with a quantization axis of the polarization of the optical fields, and the color of the arrows indicate which rf LOs couple which sublevels, and they are weighted by the Clebsch-Gordan coefficients.
}
\end{figure}

The core component of the Rydberg sensor works similarly to Ref. \cite{Meyer20}.
We use a path-stabilized optical homodyne measurement to extract EIT signals from the Rydberg sensor.
The rf heterodyne beats are then present on the optical homodyne measurement of the probe laser phase.
We can extract the amplitude and relative phases of these beats by recording a time trace of the optical homodyne signal and taking a Fast Fourier Transform (FFT).
Our EIT operates on the 5$^2$S$_{1/2} \leftrightarrow 5^2$P$_{3/2} \leftrightarrow 60^2$D$_{5/2}$ ladder in $^{85}$Rb.
The rf signal field is resonant with the $60^2$D$_{5/2} \leftrightarrow 59^2$F$_{7/2}$ transition at 10.663~GHz.

We supply our rf signal field with a single dual-polarization horn with separate inputs for two orthogonally polarized fields Sig$_h$, and Sig$_v$ as shown in Fig. \ref{fig:experimental_diagram} (a).
Both of these inputs are on resonance at 10.663~GHz, and have a maximum input power applied to the horn of -37~dBm corresponding to 140(20)~mV/m at the atoms.
At these fields the atomic response remains linear.
We can rotate the signal field polarization or change its ellipticity by changing the relative amplitude and phase of these separate input signals.
Our calibration of the input field polarization is described in Appendix~\ref{sec:signal_calibration}.
Projecting this signal field onto the cardinal axes of the LOs yields Sig$_x$, Sig$_y$ and Sig$_z$.
With the geometry of our system these projections are constrained such that $\text{Sig}_x = -\text{Sig}_h$, $\text{Sig}_y/|\text{Sig}_y| = \text{Sig}_v/|\text{Sig}_v|$ and $\text{Sig}_z/|\text{Sig}_z| = -\text{Sig}_v/|\text{Sig}_v|$.
In addition, we can manually change the polar angle ($\theta$) of the horn to vary the projection of Sig$_{v}$ into Sig$_y$ and Sig$_z$ as indicated by the grey arrow in Fig.~\ref{fig:experimental_diagram} (a).

To fully measure the polarization of an incoming rf field using our technique we require three rf LOs, one for each axis $\hat{x},\hat{y}$ and $\hat{z}$.
These LOs are applied by two identical dual-polarization rf horns, which are illustrated in Fig.~\ref{fig:experimental_diagram} a).
LO$_X$, and LO$_Y$ are applied on two orthogonal inputs of one horn, while LO$_Z$ is applied on the second horn.
Both horns are positioned so that the rf propagation is perpendicular to the optical beam paths which ensures the phase is well defined in the atomic interaction region created by the overlap of the optical beams and atomic vapor.
Each LO is at a different frequency so that all three heterodyne beats can be read out independently.
LO$_{X}$, LO$_{Y}$, and LO$_{Z}$ operate at 10.6633~GHz, 10.66327~GHz, and 10.6628~GHz respectively, which in turn are read out with beats at $\delta_X = 300$~kHz, $\delta_Y 
= 270$~kHz, and $\delta_Z = -200$~kHz.
We choose these frequencies to avoid overlapping the desired beats with harmonics, or sum and difference frequencies between the LOs.
The input rf powers for LO$_X$, LO$_Y$, and LO$_Z$ are -30.8~dBm, -31.9~dBm, and -33.1~dBm respectively, yielding approximately equal effect on the atoms as measured by the Autler-Townes splittings, and beat amplitudes.
Surrounding the cell on five sides is a box lined with rf absorbing material and  an open top for the horn inputs.
This limits reflections of the rf fields from the optics table and other metal objects in the vicinity.
However, reflections are still present, most likely from the vapor cell itself \cite{Fan15}.
These reflections are considered in Appendix~\ref{sec:rf_reflection_corrections}.

The polarization of the signal field is extracted through the amplitude and phase of the three rf heterodyne beats ($b_x$, $b_y$, and $b_z$), which are measured using an FFT of the optical homodyne output.
This FFT is taken from a 2~ms long trace with $10^6$~points.
However, since these beats are all at different frequencies there will be phase slip between them in time, and the phase difference will be dependent on the time offset of data collection.
We correct this phase slip by directly measuring the phase of LO$_Y$ and LO$_Z$ relative to LO$_X$ with an electronic mixer.
This works provided that the LOs and signal field are phase stable over the measurement time.
With phase locked synthesizers for each LO one could avoid this mixer and instead simply track these phase shifts in time, or trigger off of the greatest common denominator of the LO frequencies.

\section{\label{sec:Results} Results}

\begin{figure}[b]
\includegraphics[width = 3.375 in]{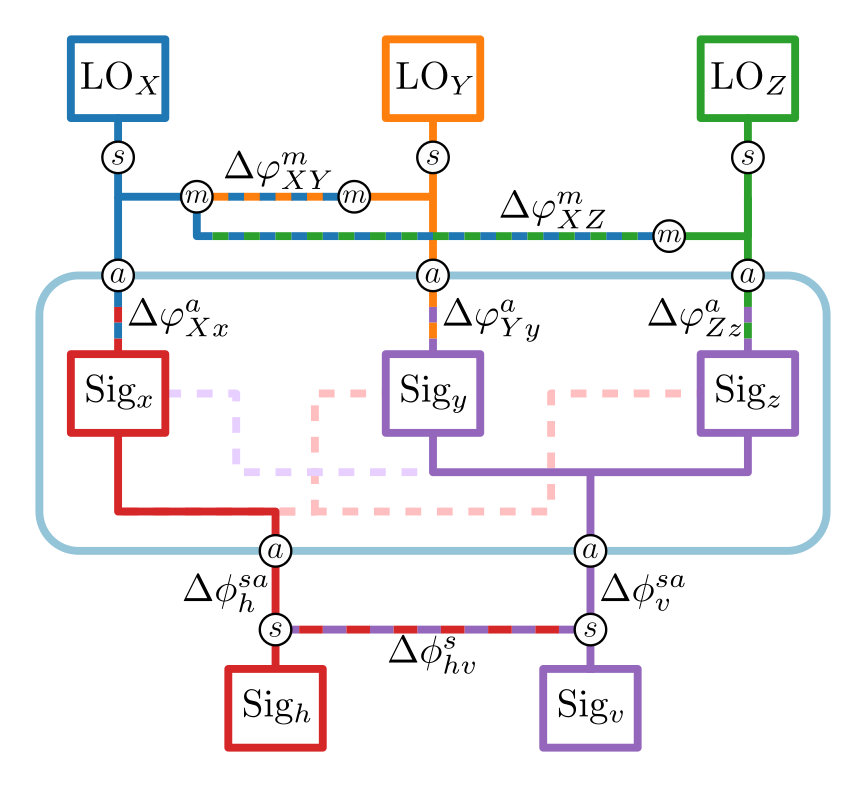}
\caption{\label{fig:phase_diagram} A graph of the phase differences in the experiment.
The multicolored dashed lines indicate phase differences that are either manually set (in the case of $\Delta\phi_{hv}^s$) or measured in the experiment.
The circled letters indicate specific locations in the experiment: each channel's synthesizer (s), the electronic mixer that mixes the LOs (m) and the atomic interaction region (a).
The phase difference between any two points on the graph can be determined by tracing a path over the graph and tracking the phase differences.
The phase differences between channels with different frequencies will change in time, while the phase differences related to propagation are constant and must be calibrated out.
}
\end{figure}

Extracting the polarization of the signal field from the output FFT of the optical homodyne measurement requires some care.
The relative amplitudes of the signal field projections along the cardinal axes can be extracted simply from the relative amplitudes of the beats measured by the optical homodyne ($|b_x|$, $|b_y|$, and $|b_z|$).
This holds provided that the LOs are properly calibrated to be of equal amplitude at the atoms.
The relative phases between the signal field projections are more complicated to obtain because they are read out at different beat frequencies.

To distinguish the phase relationships between all our different fields we introduce the notation: $\Delta \phi_{ij}^{\alpha}$, and $\Delta\phi_{i}^{\alpha \beta}$.
The indices $i$ and $j$ represent the different channels in the experiment. Namely, these are the LO and signal fields represented by their polarization axis (uppercase for LOs and lowercase for the signal).
The indices $\alpha$ and $\beta$ represent different locations in the setup: the synthesizer for each channel, the atomic interaction region, and the electronic mixer between the LOs are represented by $s$, $a$, and $m$ respectively.
Then $\Delta \phi_{ij}^{\alpha}$ is the phase difference between channels $i$ and $j$ at location $\alpha$, and $\Delta \phi_{i}^{\alpha\beta}$ is the phase difference of channel $i$ between locations $\alpha$ and $\beta$.
A graph of these phase differences is shown in Fig.~\ref{fig:phase_diagram}.
When the two channels being compared have different frequencies the phase between them will vary in time, in which case we replace $\phi$ with $\varphi$.

In our experiment we can set $\Delta \phi_{hv}^s$ by setting the phases at the synthesizer.
This sets $\Delta \phi_{hv}^a$ once the propagation delays $\Delta \phi_h^{sa}$ and $\Delta \phi_v^{sa}$ have been calibrated as described in Appendix~\ref{sec:signal_calibration}.
We extract $\Delta\phi_{hv}^a$ using our measured properties: the phases of the LO beats from the electronic mixer ($\Delta\varphi_{xy}^m$, $\Delta\varphi_{xz}^m$), and the phases of the beats from the optical signal FFT ($\Delta\varphi_{Xx}^a$, $\Delta\varphi_{Yy}^a$, $\Delta\varphi_{zZ}^a$).
Note that as LO$_Z$ is detuned below resonance, we measure $\Delta \varphi_{zZ} = -\Delta \varphi_{Zz}$ when we measure the positive frequency beat.
We have restricted Sig$_h$ to be fixed along $\hat{x}$, in effect restricting the $k$-vector to the $yz$-plane.
This allows us to extract $\Delta\phi_{hv}^a$ in two different ways, by using the phase between Sig$_x$ and either Sig$_y$ or Sig$_z$:

\begin{equation}\label{eq:phase_measurement}
\begin{split}
    \Delta\phi_{hv}^a&= \Delta \phi_{xy}^a + \pi\\
    &= -\Delta\varphi_{Xx}^a + \Delta\phi_{X}^{am} + \Delta\varphi_{XY}^m + \Delta\phi_{Y}^{ma} + \Delta\varphi_{Yy}^a + \pi\\
    &= \Delta\phi_{xz}^a\\
    &= -\Delta\varphi_{Xx}^a + \Delta\phi_{X}^{am} + \Delta\varphi_{XZ}^m + \Delta\phi_{Z}^{ma} - \Delta\varphi_{zZ}^a\\
\end{split}
\end{equation}
Note that with a $k$-vector unrestricted to the $yz$-plane, the analysis becomes more complicated and we would need to use the phases between all different signal projections.
For instance, to distinguish between left and right circular polarization with a $k$-vector along $\hat{x}$ one would have to measure $\Delta\phi^{a}_{yz}$.
Every term in Eq.~\ref{eq:phase_measurement} is either a constant phase offset due to propagation and can be calibrated out (see Appendix \ref{sec:lo_calibration}), or a phase between different channels that we can measure, and thus we can measure the signal phase difference.

\subsection{\label{sec:ResultsFixedk} Polarization measurement with a fixed $k$-vector}

First we present our measurement of polarization with a fixed $k$-vector.
For these measurements our signal horn $k$-vector was oriented with polar angle $\theta = \pi/4$, equally between $\hat{y}$ and $\hat{z}$.
In this configuration we investigated rotating a linear polarization, and changing the ellipticity.

\begin{figure*}
\includegraphics[width = \textwidth]{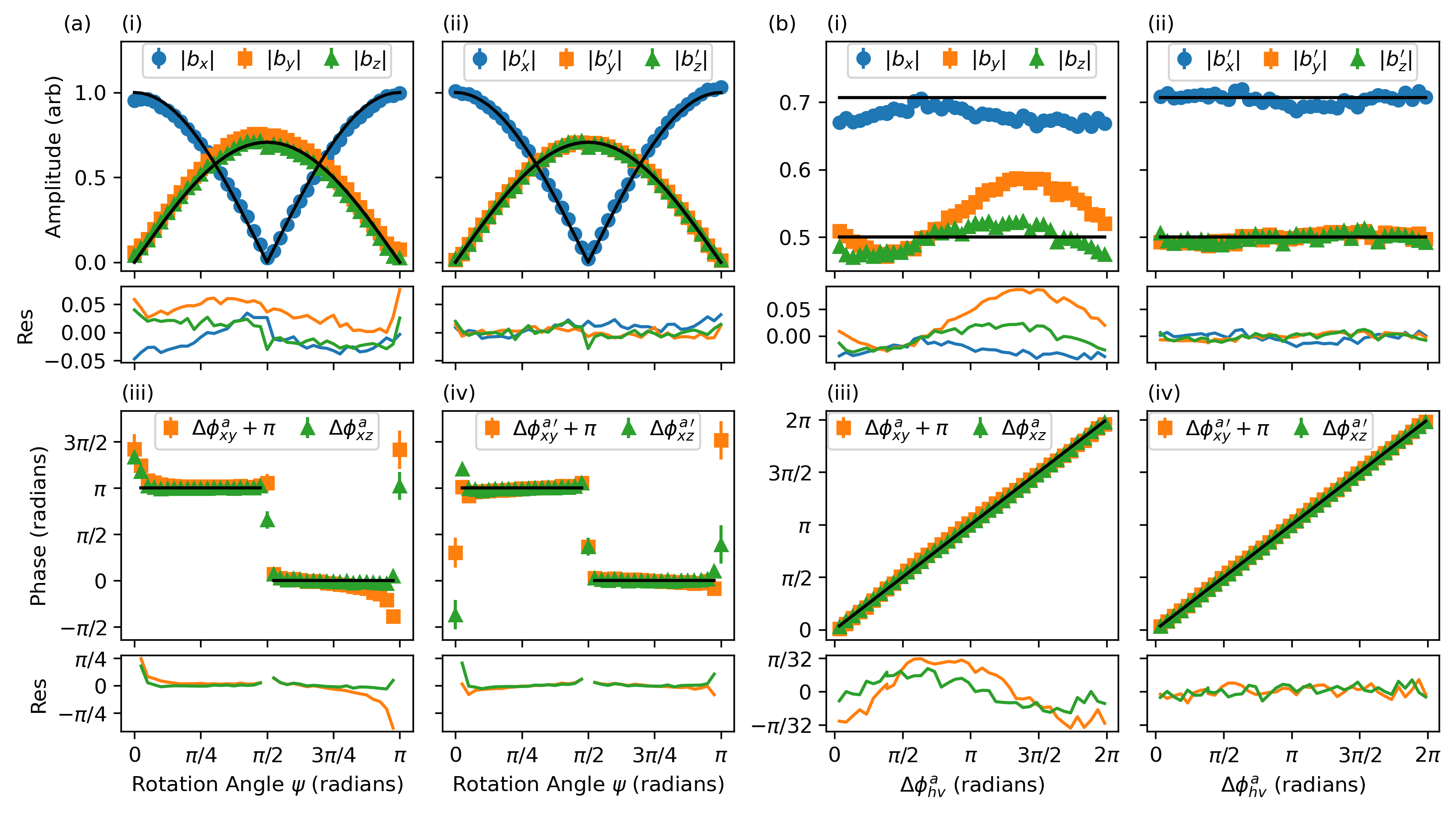}
\caption{\label{fig:rotations}
Extracted amplitudes (top row) and phases (bottom row) for linear polarization rotation (a), and changing ellipticity through the phase between Sig$_h$ and Sig$_v$ (b).
We show the data of phase offsets and normalized beat amplitude, with the LOs calibrated as in Appendix~\ref{sec:lo_calibration} ((i) and (iii)), and the result corrected for rf reflections as described in Appendix~\ref{sec:rf_reflection_corrections} ((ii) and (iv)).
The expected behavior is shown with black lines, and the residual deviations from this behavior are shown below each plot. 
}
\end{figure*}

We rotate the angle ($\psi$) of a linear signal polarization around it's $k$-vector through $\pi$ radians by changing the relative amplitudes and phases of Sig$_h$ and Sig$_v$:
\begin{equation}\label{eq:polarization_rotation}
\begin{split}
|\text{Sig}_h| \propto |\cos(\psi)|, \quad|\text{Sig}_v| \propto |\sin(\psi)|, \\
\Delta\phi_{hv}^a = \begin{cases} \pi\ if\ \psi \leq \pi/2\\
     0\ if\ \psi > \pi/2
\end{cases}
\end{split}
\end{equation}
The extracted phase differences and relative beat amplitudes from this rotation are shown in Fig.~\ref{fig:rotations} (a).
The normalized beat amplitudes and phase differences, with the LOs calibrated as in Appendix~\ref{sec:lo_calibration} without correcting for reflections are shown in Fig~\ref{fig:rotations} (a) (i) and (ii)
These largely show the expected relationship from Eq.~\ref{eq:polarization_rotation} with some deviations.
The $y$ and $z$ beat amplitudes are roughly equal as expected due to the positioning of the signal horn, but there is some offset likely due to a deviation in the LO amplitude calibration.
In addition, there is a slight horizontal shift of the beat amplitudes, as can be seen by the slight asymmetry in the points around $\psi = 0$ and $\psi = \pi/2$, and a slight rounding the the $y$ and $z$ beat amplitudes around their minimum.
The measured phase differences also have discrepancies from the expected outcome. The phase difference is undefined at $\psi = 0, \pi/2$ and $\pi$, and see some curving of the phase differences close to these values.
These discrepancies are well described by stray rf reflections, and are significantly reduced after the data is corrected as in Appendix \ref{sec:rf_reflection_corrections}.
The corrected data can seen in Fig~\ref{fig:rotations} (a) (ii) and (iv), represented by the primed beat amplitudes and phases ($|b_i'|$, $\Delta\phi^{a\prime}_{ij}$).

We also change the ellipticity of the field by altering the phase difference between Sig$_h$ and Sig$_v$ by a full 2$\pi$, while maintaining equal amplitudes.
This changes the polarization from linear along $\frac{1}{\sqrt{2}}(\hat{h} + \hat{v}) = -\frac{1}{\sqrt{2}}\hat{x} + \frac{1}{2}\hat{y} - \frac{1}{2}\hat{z}$, to right-handed circular polarization, to linear polarization along $\frac{1}{\sqrt{2}}(-\hat{h} + \hat{v}) = \frac{1}{\sqrt{2}}\hat{x} + \frac{1}{2}\hat{y} - \frac{1}{2}\hat{z}$, to left-handed circular polarization, and back.
The results of this ellipticity change are shown in Fig.~\ref{fig:rotations} (b).
Similarly to the polarization angle rotation, the uncorrected results in Fig.~\ref{fig:rotations} (b) (i) and (iii) largely match what we expect for this phase rotation with some discrepancies.
Most noticeably there is a sinusoidal variation on the beat amplitudes which should be constant.
In addition there is a small wobble on the measured phase difference.
Both these discrepancies are largely described by our reflection modeling as shown by the corrected data in Fig.~\ref{fig:rotations} (b) (ii) and (iv).

\subsection{\label{sec:ResultsPolar} Polar angle change}

\begin{figure}[b]
\includegraphics[width = 3.375 in]{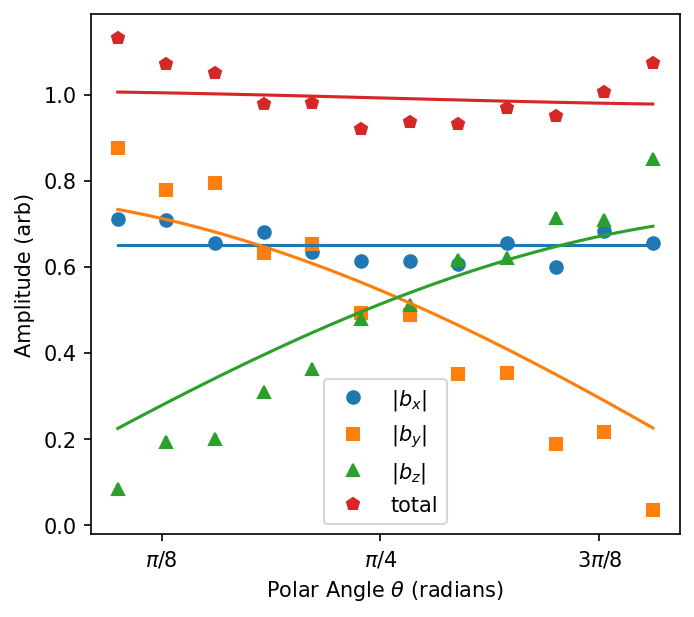}
\caption{\label{fig:PolarVariation} The beat amplitudes as the polar angle of the horn is adjusted.
The blue points are the amplitude of the $\hat{x}$ beat for polarization along $\hat{h}$, while the orange and green points are the amplitudes of the $\hat{y}$ and $\hat{z}$ beats for polarization along $\hat{v}$.
The red points are the quadrature sum of the other three.
There is more noise in these data, likely from the character of rf reflections changing as the signal horn is rotated.
The solid lines indicate the expected behavior, with a separate amplitude scaling for each beat.
}
\end{figure}

In the previous section we kept the $k$-vector of the signal field constant, restricting the signal polarizations that we could apply.
In this section, we remove this restriction. However, our ability to arbitrarily measure the polarization at any $k$-vector is limited by the the length of our interaction region along the optical axis, which is defined by the length of our vapor cell.
If the angle between the signal $k$-vector and the $k$-vector of an LO is non-zero, the phase of the heterodyne beat will vary in space.
If the extent of the interaction region along the relevant axis is larger or similar in scale to the spatial period of this phase variation the beat signal will be averaged away.
In the worst case where the $k$-vectors of the signal and LO are opposite each other this phase variation has a spatial period of twice the carrier wavelength.
Our beam widths are $\sim$400~$\mu$m, while our rf wavelengths are $\sim$3~cm, so for $k$-vectors perpendicular to the optical path this effect does not significantly influence our measurements.
However, our cell is $\sim$7.6~cm long, so this effect would come into play for $k$-vectors with a significant component along the optical axis.
Thus there is a trade-off between the size of the interaction region, the angular sensitivity dependence, and operating frequency.

We investigate our ability to detect polarizations accessible by changing the polar angle $\theta$ of our horn.
This changes the projections of Sig$_v$ along $\hat{y}$ and $\hat{z}$.
For this measurement we rotate the horn around the vapor cell in 5$^\circ$ increments and change the polarization as in Sec.~\ref{sec:ResultsFixedk}.
The results of this rotation are shown in Fig.~\ref{fig:PolarVariation}.
Since this manual rotation changes the $k$-vector of the signal horn it also changes the character of the rf reflections, this makes the data more noisy and would require a separate correction at each point to account for reflections.
Additionally, the beat amplitudes for $\hat{y}$ and $\hat{z}$ fall to zero more quickly than expected.
As our horns are only about 17~cm from our cell this could be due to the field curvature causing the beat to be partially washed out over the atomic interaction region as described above.
Additionally we could be seeing effects of variable polarization sensitivity.
These effects, in combination with reflections, could also be impacting the calibration of the LO amplitudes, which is done at a fixed $\theta = \pi/4$. 
This calibration could vary somewhat due to the different manual placements of the signal horn as the polar angle is changed, indicated by variation in the total amplitude plotted in Fig.~\ref{fig:PolarVariation}.
Regardless, the signal to noise is sufficient to clearly resolve changes in the polar angle.

\subsection{\label{sec:Constellation} Symbol reception}

\begin{figure*}
\includegraphics[width = 6.75 in]{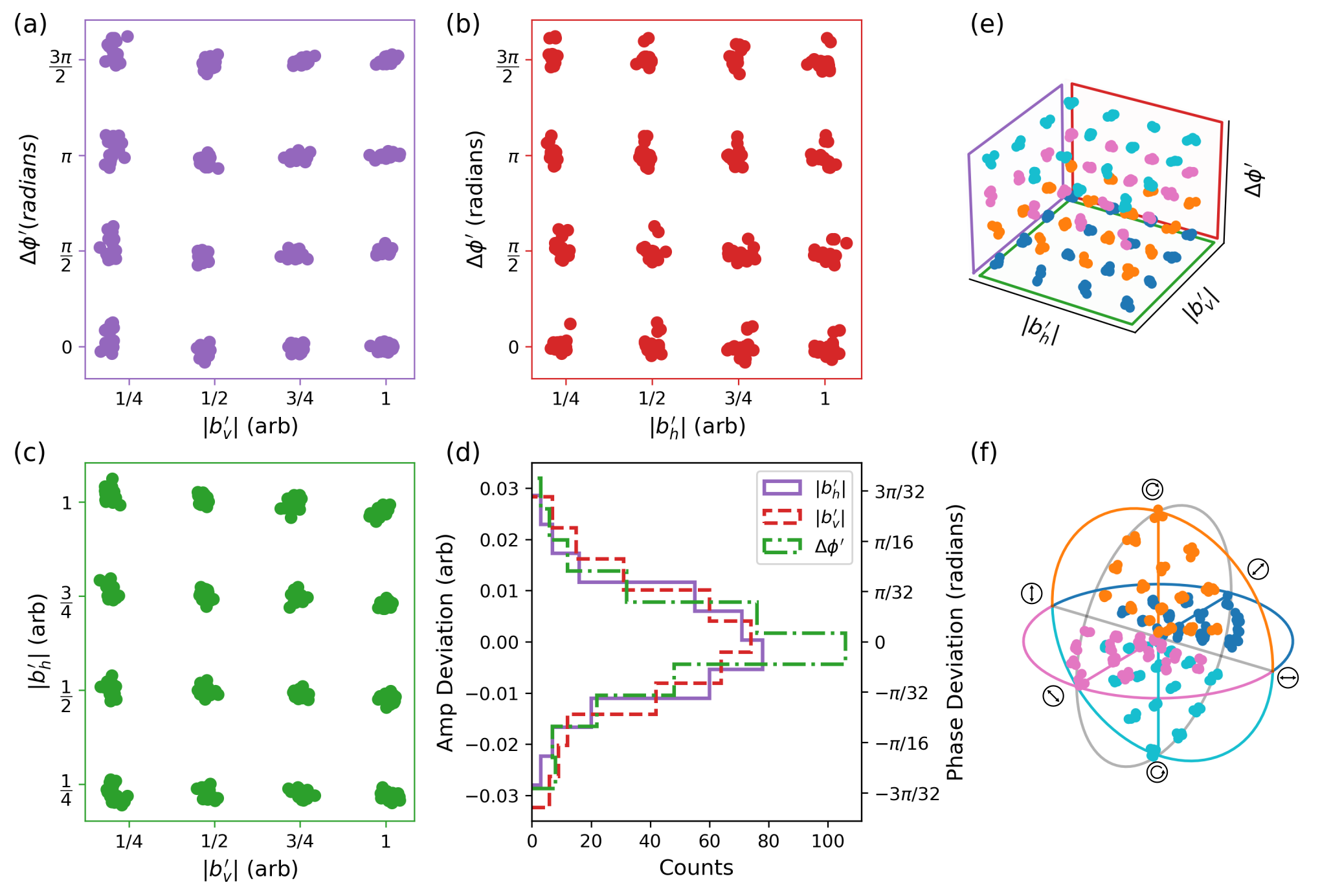}
\caption{\label{fig:Constellation}
Constellation of symbols encoded in the polarization of the signal field.
Projections of the constellation along each axis are shown in (a), (b) and (c).
These data are corrected for rf reflections.
A 3D plot of the constellation is shown in (e) with the points colored by $\Delta\phi_{hv}^{a\prime}$, and each projection plane colored to correspond with (a), (b) and (c).
A histogram of the deviations of each point from its respective symbol average along each axis.
This noise yields a volume of $0.18$~$($mV$/$m$)^2$rad per symbol, or noise density of $350$~$(\mu$V$/$m$)^2$rad/Hz when accounting for measurement time.
Lastly we plot the constellation on the poincar\'e sphere in (f).}
\end{figure*}

The added ability to measure the polarization of signals adds new degrees of freedom for symbol transmission.
We investigate our ability to distinguish data symbols encoded in the polarization space.
For a fixed $k$-vector we have three degrees of freedom available to us (excluding frequency). 
Out of convenience we encode a grid of $4^3$ symbols in the space defined by the amplitude of the field along $\hat{h}$ and $\hat{v}$, as well as the phase between them for a signal field $k$-vector as in Sec.~\ref{sec:ResultsFixedk}.
We measure the amplitude along $\hat{h}$ as $|b_h'| = |b_x'|$, the amplitude along $\hat{v}$ as $|b_v'| = \sqrt{|b_y'|^2 + |b_z'|^2}$, and the phase offset as $\Delta\phi' = \Delta\phi_{xy}^{a\prime}/2 + (\Delta\phi_{xz}^{a\prime} + \pi)/2$ the average of the two possible measurements of $\Delta \phi_{hv}^a$ using the values corrected for reflections in all three measurements.
This is not necessarily the optimal choice of encoding for symbol density, for instance one could encode in a grid inside the Poincar\'e sphere.

We transmit over all symbols 5 times and present the results of their reception in Fig.~\ref{fig:Constellation}.
Projections of the measured signal grid along each axis are shown in (a), (b), and (c).
The full grid is shown in (e), and is plotted on the Poincar\'e sphere in (f).
We also take the deviation of each point from its respective symbol mean along each axis, and plot the combined histograms over all points in (d).
the standard deviations for $|b_h'|$, $|b_v'|$, and $\Delta\phi'$ are $\sigma_h' = 1.3$~mV/m, $\sigma_v' = 1.5$~mV/m, and $\sigma_\phi' = 0.094$~rad respectively for a total volume of 0.18~$($mV$/$m$)^2$rad per symbol. 
The field amplitude standard deviations scale with the square root of the measurement time as expected yielding
noise densities of $57~\mu$V/m/$\sqrt{\text{Hz}}$, and  $66~\mu$V/m/$\sqrt{\text{Hz}}$.
In an idealized measurement the phase offset noise would similarly scale with the square root of the integration time, and be coupled to the $h$ and $v$ field amplitudes as $\sigma_\phi' = \sqrt{(\sigma_h'/|b_h'|)^2 + (\sigma_v'/|b_v'|)^2}$.
However, for our system this coupling to field amplitude is only seen at the lowest field amplitudes, such as for the leftmost points in Fig.~\ref{fig:Constellation} (a), and we do not see a strong effect from interrogation time.
This indicates that our phase measurements are limited by other effects in the system, such as non-white phase noise in the LOs or the signal itself.
When we normalize by the measurement time for the field amplitudes we get a noise density of $350$~$(\mu$V$/$m$)^2$rad/Hz.

In the analysis above we use our knowledge of the incoming $k$-vector to define the symbol basis.
Without knowledge of the incoming signal $k$-vector one would have to find the correct basis for decoding.
However, the three-dimensional nature of our measurement makes this possible.
Additionally, if we remove the restriction of a single transmitting $k$-vector and assume phase stability between all transmitters and our receiver we can expand the encoding space to all three polarization axes and their absolute phases.
This yields six degrees of freedom total for a single carrier frequency.

\section{Conclusions}
We demonstrate three-dimensional polarization measurements with a Rydberg electrometer using rf heterodyne detection to measure the projections of the field along each cardinal axis, and their relative phases.
This allows us to not only measure a linear polarization angle, but also the ellipticity of the field.
These measurements are done simultaneously without interfering with the system's ability to measure the total field magnitude.
We illustrate these measurements with a fixed signal field $k$-vector, and when the $k$-vector is rotated in space.
Lastly we also investigate the ability to receive information encoded in the polarization and amplitude of a signal.

This system could measure the $k$-vector of an incoming field by measuring the plane of the polarization ellipse. 
However, the ability of our system to sense from arbitrary $k$-vectors is limited by the atomic interaction region being longer than the rf wavelength, causing signal washout for some $k$-vectors.
This is not a fundamental limitation, and could be resolved with a symmetric and appropriately sized interaction region.
Alternatively, one could include additional LOs with redundant polarizations but orthogonal $k$-vectors.
This would improve the ability to detect signals from all directions, even with a larger interaction region, and could give $k$-vector information for linearly polarized fields by looking at the relative strength of the beats with each redundant LO.

\begin{acknowledgments}
We thank Fredrik Fatemi for valuable discussions.
The views, opinions and/or findings expressed are those of the authors and should not be interpreted as representing the official views or policies of the Department of Defense or the U.S. Government.
References to commercial devices do not constitute an endorsement by the U.S. Government or the Army Research Laboratory.
They are provided in the interest of completeness and reproducibility.
\end{acknowledgments}

\appendix

\section{Phase and amplitude calibrations of the signal \label{sec:signal_calibration}}
As we are transmitting our own signal field we need to properly calibrate the angle and ellipticity of this field in order to evaluate the accuracy of our Rydberg atom polarimeter.
While we can independently set the phase of Sig$_h$ and Sig$_v$ at their respective synthesizers, we must ensure the absolute phase difference at the atoms $\Delta\phi_{hv}^a$ is known.
There are two effects that make this difficult.
First, there are propagation differences between Sig$_h$ and Sig$_v$ as they travel from the synthesizers to the horn, and in the phase of their clocking signals.
Second, we found that there was an output power dependent phase slip in the output of the Windfreak SynthHD PRO synthesizers used for the signal field.
We measure this second effect separately with a mixer and are able to correct for it when setting the phase difference, however some effect may remain and explain some of the residual discrepancies in the phases of Fig.~\ref{fig:rotations} (a) (ii) and (iv).
To account for a fixed phase offset between the signal fields requires a different approach.
With the horn aligned as in Sec.~\ref{sec:ResultsFixedk} each LO only beats with either Sig$_h$ or Sig$_v$.
This means that a phase offset in the signal fields is indistinguishable from a phase offset in the LOs.
Instead, to calibrate this phase offset we rotate the horn by $\pi/4$ around it's $k$-vector so that $\hat{h} = -\frac{1}{\sqrt{2}}\hat{x} + \frac{1}{2}\hat{y} - \frac{1}{2}\hat{z}$, and $\hat{v} = \frac{1}{\sqrt{2}}\hat{x} + \frac{1}{2}\hat{y} - \frac{1}{2}\hat{z}$.
When we attempt to rotate a linear polarization in this configuration any unaccounted for phase difference between the signal fields will cause some ellipticity change.
This will show up as a linear slope to the measured phase differences of the beats.
Thus by adjusting this slope to zero we can zero the residual phase difference.
This calibrates our signal field so that we can accurately set linear or circular polarization at any angle.
Using this method we find a phase offset of 1.2~radians that must be accounted for when setting the phase offset at the synthesizers.

Amplitude differences caused by different gain between the two signal channels can be calibrated independently from the LO amplitdues in a similar way.
We rotate the polarization while the signal horn is physically rotated by $\pi/4$.
If the signal fields have equal amplitude the peak beat amplitude for $\hat{x}$ will be separated from the peak beat amplitudes for $\hat{y}$ and $\hat{z}$ by $\pi/2$.
However, differences between the maximum $|\text{Sig}_h|$ and $|\text{Sig}_v|$ will shift this offset.
Thus we correct for any signal amplitude asymmetry by adjusting the difference between the peak amplitudes to $\pi/2$.
We find an amplitude ratio of 1.1 that must be accounted for when setting the amplitudes at the synthesizers.

\section{Phase and amplitude calibrations of the LOs \label{sec:lo_calibration}}

In order to correctly compare the phase and amplitude of the rf heterodyne beats we must have relative phase and amplitude information about the LOs.
As described in the main text, by measuring the relative phases of the LOs with an electronic mixer we can correct for phase slips due to the frequency difference of the LOs.
This allows us to accurately detect changes in the phase difference $\Delta\phi_{hv}^a$ from shot to shot, but does not provide an absolute measure.
In order to attain an absolute measurement we must account for the constant phase offsets in the LOs due to propagation delays between the mixer and the atoms, notably $\Delta\phi_x^{ma}$, $\Delta\phi_y^{ma}$, and $\Delta\phi_z^{ma}$.
We correct for these delays by offsetting $\Delta\phi_{xy}^a + \pi$ and $\Delta\phi_{xz}^a$ to $0$ when the signal field is linear along $1/\sqrt{2}\hat{x} + 1/2\hat{y} + 1/2\hat{z}$ after signal calibration.
Similarly, we calibrate the LO amplitudes by adjusting them until the beat amplitudes are correctly related as $|b_y| = |b_z|$, and $|b_x|^2 = |b_y|^2 + |b_z|^2$.
Alternatively we can make these calibrations through our rf reflection calibration described in Sec. \ref{sec:rf_reflection_corrections}.
In a more highly engineered system it would likely be possible to correctly calibrate these phase and amplitude offsets through direct measurement rather than relying on calibrating with a known signal field.
In addition, it could be possible to directly measure the LO phases with the atoms themselves through higher order terms of Eq. \ref{eq:at_splitting}.

\section{rf reflection corrections \label{sec:rf_reflection_corrections}}

\begin{figure}
\includegraphics[width = 3.375 in]{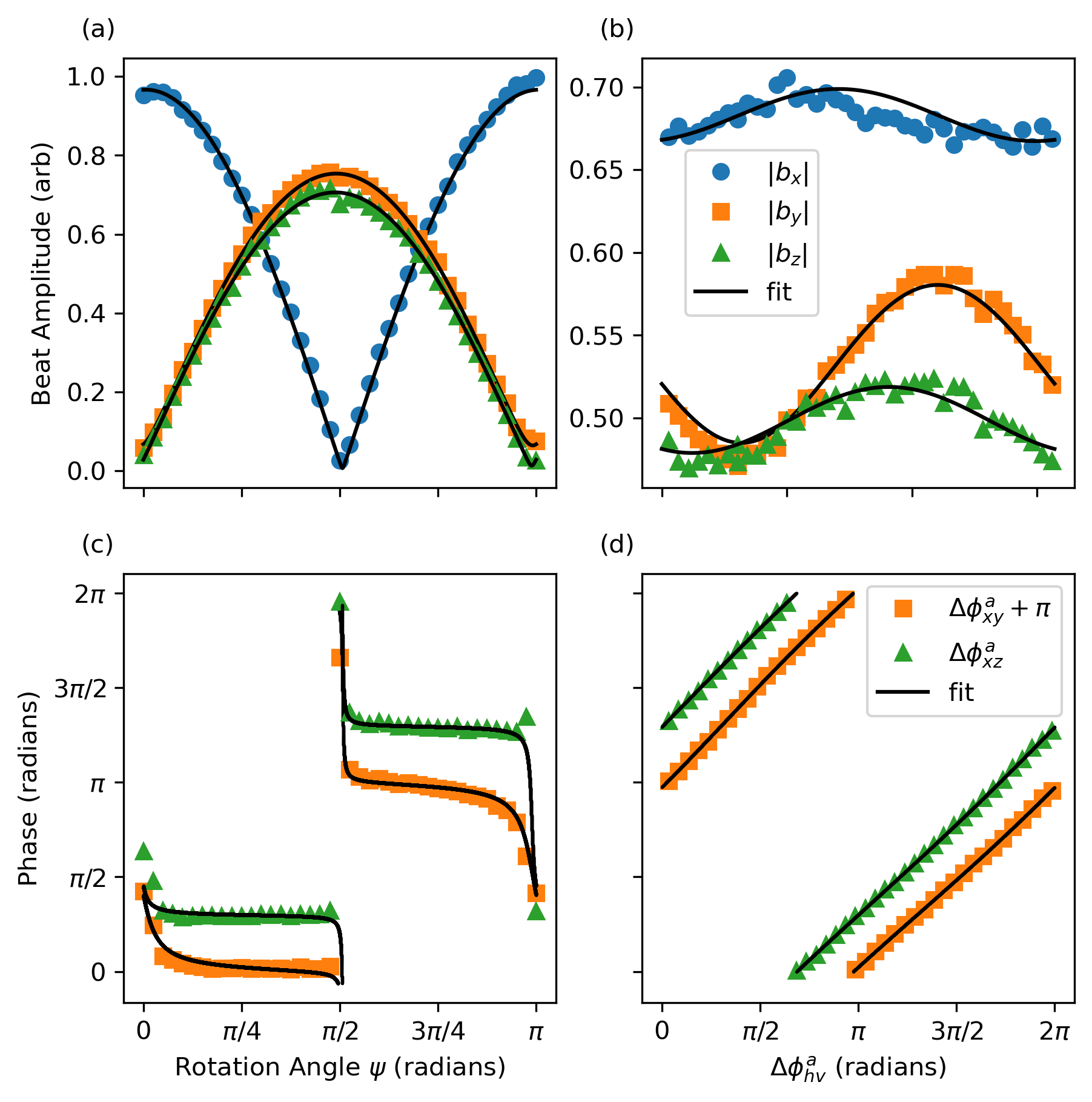}
\caption{\label{fig:reflection_fitting} Modeling of the effect of rf reflections in the system.
(a) Fitted model to beat amplitudes from Fig. \ref{fig:rotations} (a) (i).
(b) Fitted model to beat amplitudes from Fig. \ref{fig:rotations} (b) (i).
(c) Fitted model to phase measurements from Fig. \ref{fig:rotations} (a) (iii).
(d) Fitted model to phase measurements from Fig. \ref{fig:rotations} (b) (iii).
All these fits were performed simultaneously.
The phase calibration described in Appendix \ref{sec:lo_calibration} was not performed for these data}
\end{figure}

While we have attempted to eliminate most reflections of the rf fields from the optical table and other components, some still remain.
These are largely from the vapor cell walls, and their effect can be seen in discrepancies between our data and the expected output.
However, as long as these reflections are constant we can use a simple model to analyze and correct for them.
Without reflections only the projection of the signal along $\hat{x}$ and LO$_X$ should impact $b_x$ to first order.
However, if some portion of LO$_X$ reflects into $\hat{y}$ then the projection of the signal along $\hat{y}$ will also have an impact on this beat.
To model these reflections we introduce the reflection matrix:
\begin{equation}
A =  \begin{bmatrix}
    A_{xx}& A_{xy} & A_{xz}\\
    A_{yx} & A_{yy} & A_{yz}\\
    A_{zx} & A_{zy} & A_{zz}
    \end{bmatrix},
\end{equation}
where the elements $A_{ij}$ are complex numbers that account for the amplitude and phase offset of the combined field of LO$_i$ and all of it's reflections projected onto $\hat{j}$.
While we frame this matrix as modeling reflections solely from the LOs, we cannot distinguish between LO reflections and signal reflections.
This does not affect our ability to correct for reflections, but it does mean the resulting $A$ matrix does not fully describe which fields are reflecting.
In the ideal case where all LOs have an equal amplitude normalized to 1, no phase offsets (relative to the electronic mixer), and no reflections this matrix becomes the identity matrix.
In the case that there are no reflections, but the LOs are not properly calibrated this model can also account for the corrections in Appendix.~\ref{sec:lo_calibration}; any residual differences in the LO amplitudes and LO phase offsets will appear on the diagonal.
With a fully populated matrix accounting for all reflections the resulting beats are 
\begin{equation}
    \begin{bmatrix}
        b_x\\
        b_y\\
        b_z^*
    \end{bmatrix} = A \cdot \begin{bmatrix}
        Sig_x\\
        Sig_y\\
        Sig_z
    \end{bmatrix}^*.
\end{equation}
Note that the complex conjugation of $b_z$ is due to the negative detuning of LO$_Z$.
We use this model to simultaneously fit all the data from Sec. \ref{sec:ResultsFixedk} to a single $A$ matrix.

The results of this fitting are shown in Fig. \ref{fig:reflection_fitting}, and yield a matrix
\begin{equation}
A = \begin{bmatrix}
    0.966               & 0.017 e^{0.72 i}  & 0.017 e^{3.05 i} \\
    0.067 e^{4.41 i}   & 1.057 e^{3.15 i}  & 0.012 e^{5.48 i} \\
    0.028 e^{1.42 i}    & 0.021 e^{1.86 i}  & 0.985 e^{4.06 i}
\end{bmatrix},
\end{equation}
where all phases are referenced to $A_{xx}$.
For our data this result in not fully unique as $\text{Sig}_y = \text{Sig}_z$ but regardless it can be used to correct for the reflections present in our system by applying the reverse transformation:
\begin{equation}
    \begin{bmatrix}
        b_x'\\
        b_y'\\
        b_z^{*\prime}
    \end{bmatrix} = \left(A^{-1}\right)^* \cdot \begin{bmatrix}
        b_x\\
        b_y\\
        b_z
    \end{bmatrix}^*.
\end{equation}
This model could also be accounting for some higher order terms in Eq. \ref{eq:at_splitting}.


\bibliography{polarimeter}

\end{document}